\documentclass[a4paper,12pt]{article}
\usepackage[dvips,colorlinks=true,linkcolor=blue,urlcolor=red]{hyperref}
\usepackage{amsmath}%
\usepackage{amsfonts}%
\usepackage{amsthm} 
\newcommand{\be}{\begin{equation}}
\newcommand{\ee}{\end{equation}}
\newcommand{\one}{\mathbf{1}}
\newcommand{\lap}{\bigtriangleup}

\newcommand{\C}{\mathbb C}
\newcommand{\R}{\mathbb R}
\newcommand{\N}{\mathbb N}

\newcommand{\rkl}{\rangle}
\newcommand{\lkl}{\langle}
\newcommand{\D}{\mathcal{D}}

\newcommand{\cS}{\mathcal{S}}
  
\newcommand{\tr}{\mathrm{Trace}}  
\newcommand{\supp}{\mathrm{supp}}

\newcommand{\cT}{\mathcal{T}}

\newcommand{\bbE}{\mathbb E}

\newcommand{\ol}{\overline}
\newcommand{\ul}{\underline}

\newtheorem{theorem}{Theorem}[section]        
\newtheorem{lemma}[theorem]{Lemma}            
\newtheorem{corollary}[theorem]{Corollary}

\theoremstyle{plain}
\begin{document}
\title{The integrated density of states for an interacting
  multielectron homogeneous model}
\author{Fr{\'e}d{\'e}ric Klopp \\ LAGA, Institut Galil{\'e}e; Universit{\'e}
  Paris-Nord; \\
  F-93430 Villetaneuse; France \\
  {\em email:}
  \href{mailto:klopp@math.univ-paris13.fr}{klopp@math.univ-paris13.fr}
  \and
  Heribert Zenk\\ Mathematisches Institut,
  Ludwig-Maximilians-Universit{\"a}t\\
  Theresienstra{\ss}e 39, 80333 M{\"u}nchen \\
  {\em email:}
  \href{mailto:Heribert.Zenk@mathematik.uni-muenchen.de}{Heribert.Zenk@mathematik.uni-muenchen.de}
}
\maketitle
\begin{abstract}
  \noindent For a system of $n$ interacting electrons moving in the
  background of a ``homogeneous'' potential, we show that, if the
  single particle Hamiltonian admits a density of states, so does the
  interacting Hamiltonian. Moreover this integrated density of states
  coincides with that of the free electron Hamiltonian.
\end{abstract}

\section{Introduction}
\setcounter{equation}{0}
We consider $n$ interacting electrons moving in a ``homogeneous'' electric
field in the $d$-dimensional configuration space $\R^d$. A typical
example of what we mean by a ``homogeneous'' potential is an Anderson
or alloy-type random potential. The goal of the present short note is
to prove that, if the Hamiltonian of the single particle in the
``homogeneous'' media admits an integrated density of states (IDS),
then, so does the interacting $n$-particle Hamiltonian. Moreover, this
IDS is equal to that of the interacting $n$-particle Hamiltonian.
Heuristically, this is easily understood : it follows from the fact
that the electron-electron interaction essentially lives on a strict
sub-manifold of the total configuration space, whereas the IDS is a
thermodynamic limit over the whole space and, as such, measures
``bulk'' phenomena.
\subsection{The interacting multi-electron model}
\label{sec:inter-mult-model}
The non-interacting $n$-electron Hamiltonian satisfies $H_n=-\lap
+V_n$ where the Laplacian $-\lap$ on $\R^{nd}$ describes the free
kinetic energy of the $n$ electrons. As all the electrons are in the
same background, the potential $V_n$ is of the form
\begin{equation} \label{eq1.3}
  V_n=\sum_{k=1}^n \one_{L^2(\R^d)}^{\otimes (k-1)}
  \otimes V_1 \otimes
  \one_{L^2(\R^d)}^{\otimes (n-k)}.
\end{equation}
Hence, the noninteracting $n$-electron Hamiltonian takes the form
\begin{equation} 
  \label{gl4.1}
  H_n:=\sum_{k=1}^n \one_{L^2(\R^d)}^{\otimes (k-1)} \otimes 
  H_1 \otimes \one_{L^2(\R^d)}^{\otimes (n-k)}\text{ where
  }H_1=-\Delta+V_1.
\end{equation} 
On the one particle potential $V_1$, we assume that
\begin{description}
\item[(H.1.a)] $(V_1)_+:=\max \{V_1,0\}$ is locally square integrable
  and $(V_1)_-:=\max\{-V_1,0\}$ is an infinitesimally $-\lap$-bounded
  potential that is $\D((V_1)_-)\supseteq\D(-\lap)$ and for all
  $\alpha >0$, there exists $\gamma(\alpha)<\infty$, such that for all
  $\phi\in\D(-\lap)$
\begin{equation} 
  \label{glH1.a}
  \|(V_1)_-\phi\|\leq\alpha\|\lap\phi\|+\gamma(\alpha)\|\phi\| 
\end{equation}
\item[(H.1.b)] the operator $H_1$ admits an integrated density of
  states, say $N_1$, that is, if $H_{1|\Lambda(0,L)}$ denotes the
  Dirichlet restriction of $H_1$ to a cube $\Lambda(0,L)$ centered at
  0 of side-length, $L$, then the following limit exists
  \begin{equation*}
    N_1(E):=\lim_{L\to+\infty}L^{-d}\tr(\one_{]-\infty,E]}
    (H_{1|\Lambda(0,L)})).  
  \end{equation*}
\end{description}
Assumption (H.1.a) implies essential self-adjointness of $-\lap+V_1$ on
$C_0^{\infty}(\R^d)$ by \cite{RS2} Theorem X.29. Indeed,
\begin{equation}
  \label{eq:2}
  V_n=V_n^+-V_n^-,\quad
  V_n^\pm(x_1,...,x_n):= \sum_{j=1}^n (V_1)_\pm(x_j)
\end{equation}
where
\begin{itemize}
\item $V_n^-$ is infinitesimally $-\lap$-bounded i.e.~\eqref{glH1.a}
  holds for the same constants and the Laplacian over $\R^{nd}$,
\item $V_n^+$ is non-negative locally square integrable.
\end{itemize}
The self-adjoint extensions of $-\lap+V_1$ and $-\lap+V_n$ are again
denoted by $H_1$ and $H_n$; they are bounded from below.\\
Classical models for which the IDS is known to exist include periodic,
quasi-periodic and ergodic random Schr{\"o}dinger operators (see
e.g.~\cite{PaFi}).\\
In the definition of the density of states, we could also have
considered the case of Neumann or other boundary conditions.
\vskip.2cm The interacting $n$-particle Hamiltonian is of the form
\begin{equation} \label{eq1.1}
H:=-\lap+V_i+V_n, 
\end{equation} 
where
\begin{equation} \label{eq1.2b}
V_i(x_1,...,x_n):=\sum_{1 \leq k < l \leq n} V(x_k-x_l)
\end{equation} 
is a localized repulsive interaction potential generated by the
electrons; so we assume
\begin{description}
\item[(H.2)] $V:\ \R^d\to\R$ is measurable non-negative, locally
  square integrable and $V$ tends to $0$ at infinity.
\end{description}
\vskip.2cm\noindent The standard repulsive interaction in
three dimensional space is of course the Coulomb interaction
$V(x)=1/|x|$. In some cases, due to screening, it must be replaced by
the Yukawa interaction $V(x)=e^{-|x|}/|x|$.
\vskip.2cm Finally, we make one more assumption on both $V_1$ and $V$
: we assume that
\begin{description}
\item[(H.3)] the operator $V_i(H_n-i)^{-1}$ is bounded.
\end{description}
Assumption (H.3) is satisfied in the case of the Coulomb and Yukawa
potential for those $V_1$ satisfying (H.1.a): $H_n$ is self-adjoint on
$\D(H_n) \subseteq \D(-\lap)$, hence $\|V_i(H_n-i)^{-1}\| \leq \|V_i
(-\lap-i)^{-1} \| \cdot \|(-\lap-i)(H_n-i)^{-1}\|$, where $
\|(-\lap-i)(H_n-i)^{-1} \| < \infty$ due to closed graph theorem and $
\|V_i (-\lap-i)^{-1} \| < \infty$ for Coulomb and Yukawa interaction
potentials $V_i$, see \cite{RS2} Theorem X.16.
\section{The integrated density of states} 
\label{kap4}
\setcounter{equation}{0}
We now compute the IDS for the $n$-electron model. Let $\Lambda_L =
\Lambda(0,L)$ be the cube in $\R^d$ centered at $0$ with side-length
$L$ and write $\Lambda_L^n=\Lambda_L\times\cdots\times \Lambda_L$ for
the product of $n$ copies of $\Lambda_L$. We denote the restriction of
the interacting $n$-particle Hamiltonian $H$ to $\Lambda_L^n$ with
Dirichlet boundary conditions by $H_{|\Lambda_L^n}$.  Clearly
assumptions (H.2) and (H.1.a) guarantee that $H_{|\Lambda_L^n}$ is
bounded from below with compact resolvent. Hence, for any $E\in\R$,
one defines the normalized counting functions
\begin{equation*}
  N_L(E):=L^{-nd}\tr(\one_{]-\infty,E]}(H_{|\Lambda_L^n})).
\end{equation*}
As usual, $N$, the IDS of $H$ is defined as the limit of $N_L(E)$ when
$L\to+\infty$. Equivalently, one can define the density of states
measure applied to a test function $\varphi$ as the limit of
$L^{-nd}\tr[\varphi( H_{|\Lambda_L^n})]$. If the limit exists, it
defines a non-negative measure. It is a classical result that the
existence of that limit (for all test functions) or that of $N_L(E)$
are equivalent (\cite{PaFi}).
\subsection{The IDS for the noninteracting
  $n$-electron system}
\label{sec:integr-dens-stat}
Recall that, by assumption (H.1.b), the single particle model $H_1$
admits an IDS (see~\cite{PaFi}) and a density of states measure
denoted respectively by $N_1$ and $\nu_1$.\\
Let $H_{n|\Lambda_L^n}$ be the restriction of $H_n$ to $\Lambda_L^n$
with Dirichlet boundary conditions. One has
\begin{lemma} 
  \label{lemma4.1}
  The IDS for the noninteracting $n$-electron model given by
  \begin{equation} 
    \label{gl4.2}
    N_n(E):= \lim_{L \to \infty} \frac{1}{L^{nd}} \tr
    (\one_{]-\infty,E]} H_{n|\Lambda_L^n})
  \end{equation} 
  exists and satisfies
  \begin{equation} 
    \label{gl4.2a}
    N_n=N_1 \ast \nu_1 \ast \cdots \ast \nu_1.
  \end{equation} 
\end{lemma}
\noindent Let us comment on this result. First, the convolution product
in~(\ref{gl4.2a}) makes sense as all the measures and functions are
supported on half-axes of the form $[a,+\infty)$; this results from
assumption (H.1.a) . When the field $V_1$ is not bounded from below,
one will need some estimate on the decay of $N_1$ and $\nu_1$ near
$-\infty$ to make sense of~(\ref{gl4.2a}) (and to prove it); such
estimates are known for some models (see e.g.~\cite{PaFi,KlPa}).
\vskip.2cm\noindent
{\it Proof of Lemma~\ref{lemma4.1}.} The operator $H_n$ is the sum of
$n$ commuting Hamiltonians, each of which is unitarily equivalent to
$H_1$; so is $H_{n|\Lambda_L^n}$, its restriction to the cube
$\Lambda_L^n$. As the sum decomposition of $H_n$ commutes with the
restriction to $\Lambda_L^n$, the eigenvalues of $H_{n|\Lambda_L^n}$
are exactly the sum of $n$ eigenvalues of $H_1$ restricted to
$\Lambda_L$. This immediately yields that
  \begin{equation*}
    \tr(\one_{]-\infty,E]}(H_{n|\Lambda_L^n}))=(\hat
    N_1^L*\hat\nu_1^L*\cdots*\hat\nu_1^L)(E)
  \end{equation*}
  where $\hat N_1^L(E)$ is the eigenvalue counting function for $H_1$
  restricted to $\Lambda_L$, and $\hat\nu_1^L$ its counting measure
  (i.e. $d\hat N_1^L$). The normalized counting function and measure,
  $N_1^L$ and $\nu_1^L$, are defined as
  \begin{equation*}
    N_1^L=\frac{1}{L^d}\hat N_1^L\quad\text{ and
    }\quad\nu_1^L=\frac{1}{L^d}\hat\nu_1^L.
  \end{equation*}
  The existence of the density of states of $H_1$ then exactly says
  that $N_1^L$ and $\nu_1^L$ converge respectively to $N_1$ and
  $\nu_1$. The convergence of $N_1^L*\nu_1^L*\cdots*\nu_1^L$ to
  $N_1*\nu_1*\cdots*\nu_1$ is then guaranteed as the convolution is
  bilinear bi-continuous operation on distributions. This completes
  the proof of Lemma~\ref{lemma4.1}.\qed\vskip.2cm
Let us now say a word on the boundary conditions chosen to define the
IDS. Here, we chose to define it as a thermodynamic limit of the
normalized counting for Dirichlet eigenvalues. Clearly, if we know
that the single particle Hamiltonian has a IDS defined as the
thermodynamic limit of the normalized counting for Neumann
eigenvalues, so does the non-interacting $n$-body Hamiltonian.
Moreover, in the case when the two limits coincide for the one-body
Hamiltonian, they also coincide for the non-interacting $n$-body
Hamiltonian. Using Dirichlet-Neumann bracketing, one then sees that
the integrated densities of states for both the one-body and
non-interacting $n$-body Hamiltonian for positive mixed boundary
conditions also exist and coincide with that defined with either
Dirichlet or Neumann boundary conditions.
\subsection{Existence of the IDS for the interacting $n$-electron system}
\label{sec:exist-integr-dens}
Let $H_{|\Lambda_L^n}$ denote the restriction of $H$ to the box
$\Lambda_L^n$ with Dirichlet boundary conditions. Our main result is
\begin{theorem} 
  \label{th4.2}
  Assume (H.1), (H.2) and (H.3) are satisfied. For any $\varphi \in
  C_0^{\infty}(\R)$, one has
  \begin{equation} 
    \label{gl4.11}
    \frac{1}{L^{nd}}  \tr
    [\varphi(H_{|\Lambda_L^n})-\varphi(H_{n|\Lambda_L^n}
    )] \stackrel {L \to \infty}{\longrightarrow} 0. 
  \end{equation} 
\end{theorem}
\noindent As the density of states measure of $H$ is defined by
\begin{equation*}
  \langle\varphi,dN\rangle=\lim_{L\to+\infty}\frac{1}{L^{nd}}\tr[
    \varphi(H_{|\Lambda_L^n})].
\end{equation*}
we immediately get the
\begin{corollary}
  \label{cor4.3}
  Assume (H.1), (H.2) and (H.3) are satisfied. The IDS for the
  interacting $n$-electron model $H$ exists and coincides with that of
  the noninteracting model $H_0$; hence, it satisfies
  \begin{equation*}
    N=N_n=N_1 \ast \nu_1 \ast \cdots \ast \nu_1.    
  \end{equation*}
\end{corollary}
Note that, in view of the remark concluding
section~\ref{sec:integr-dens-stat}, we see that the integrated density
of states of the interacting $n$-body Hamiltonian is independent of
the boundary conditions if that of the one-body Hamiltonian is.
\par One of the interesting properties of the integrated density of
states is its regularity; it is well known to play an important role
in the theory of localization for random one-particle models (see
e.g.~\cite{Stl}). Usually, it comes into play through a Wegner
estimate i.e. an estimate of the type
\begin{equation*}
  \bbE (\tr \one_{]E_0-\eta,E_0+\eta[}(H_{|\Lambda_L^n})) \leq 
  C_W \eta |\Lambda_L^n|
\end{equation*}
For a specific model of random one-particle Hamiltonian, a Wegner
estimate for the IDS of the interacting Hamiltonian was proved
in~\cite{Ze2}. This estimate yields Lipshitz continuity of the IDS for
that model.
\par On the other hand, Corollary~\ref{cor4.3} directly relates the
regularity of the IDS of the interacting system to that of the IDS of
the single particle Hamiltonian. The regularity of the IDS of the
single particle has been the subject of a lot of interest recently
(see e.g.~\cite{CHK,Stz}).
\vskip.2cm\noindent
{\it Proof of Theorem~\ref{th4.2}.} We take some $q > \frac{nd}{2}$ and
specify the appropriate choice later on. By assumption (H.1.a) and
(H.2), there exists $\zeta>0$ such that
  \begin{equation} 
    \label{gl4.11a} 
    -\infty<-\zeta\leq\min\left(\inf_{L\geq1}\{\inf[\sigma
      (H_{n|\Lambda_L^n})\cup\sigma(H_{|\Lambda_L^n})]\},
      \inf[\sigma(H_n)\cup\sigma(H)]\right).
  \end{equation} 
  Let $\gamma=\gamma(1/2)$ be given by~\eqref{glH1.a} for
  $\alpha=1/2$. Fix $\lambda_0>\zeta+2\gamma+1$.\\
  By~\eqref{gl4.11a}, we only need to prove~\eqref{gl4.11} for
  $\varphi\in C_0^{\infty}(\R)$ supported in $(-\zeta-1,+\infty)$. For
  such a function, let $\tilde{\varphi}$ be an almost analytic
  extension of the function $x\mapsto(x+\lambda_0)^q\varphi(x)\in
  C_0^{\infty}(\R)$ i.e.  $\tilde{\varphi}$ satisfies
  \begin{itemize}
  \item $\tilde\varphi\in \cS(\{z\in\C: |\Im z|<1\}$
  \item for any $k\in\N$, the family of functions
    $(x\mapsto\frac{\partial \tilde{\varphi}}{\partial
      \ol{z}}(x+iy)|y|^{-k})_{0<|y|<1}$ is bounded in $\cS(\R)$.
  \end{itemize}
  The functional calculus based on the Helffer-Sj{\"o}strand formula
  implies
  \begin{eqnarray} 
    \lefteqn{\varphi(H_{|\Lambda_L^n})-\varphi(H_{n|\Lambda_L^n} )=
      \label{gl4.12}}\\
    &=&\frac{i}{2\pi} \int_{\C} \frac{\partial
      \tilde{\varphi}}{\partial{\ol{z}}}(z)
    [(H_{|\Lambda_L^n}+\lambda_0)^{-q}(H_{|\Lambda_L^n}-z)^{-1}-
    \nonumber\\
    && \hspace{2.5cm}
    (H_{n|\Lambda_L^n}+\lambda_0)^{-q}(H_{n|\Lambda_L^n}-z)^{-1}]
    dz \wedge d\ol{z}. \nonumber
  \end{eqnarray}
  In the following, we apply an idea, which has already been used in
  \cite{Kl} and \cite{KlPa} and which simplifies in this situation.
  Using resolvent equality, the integrand in (\ref{gl4.12}) is written
  as
  \begin{equation}
    \label{gl4.17}
    \begin{split}
      (H_{|\Lambda_L^n}+\lambda_0)^{-q}&(H_{|\Lambda_L^n}-z)^{-1}-
      (H_{n|\Lambda_L^n}+\lambda_0)^{-q}(H_{n|\Lambda_L^n}-z)^{-1}=\\
      &=(H_{n|\Lambda_L^n}+\lambda_0)^{-q}
      [(H_{|\Lambda_L^n}-z)^{-1}-(H_{n|\Lambda_L^n}-z)^{-1}]\\
      &\hskip3cm+[ (H_{|\Lambda_L^n}+\lambda_0)^{-q}-
      (H_{n|\Lambda_L^n}+\lambda_0)^{-q}](H_{|\Lambda_L^n}-z)^{-1} \\
      &= -(H_{n|\Lambda_L^n}+\lambda_0)^{-q}
      (H_{n|\Lambda_L^n} -z)^{-1} V_i (H_{|\Lambda_L^n}-z)^{-1}\\
      &\hskip3cm- \sum_{l=1}^q (H_{n|\Lambda_L^n}+\lambda_0)^{l-q-1}
      V_i (H_{|\Lambda_L^n}+\lambda_0)^{-l}(H_{|\Lambda_L^n}-z)^{-1}
    \end{split}
  \end{equation}
  Estimating the trace of (\ref{gl4.17}), we choose $\varepsilon >0$
  and write
  \begin{equation} 
    \label{gl4.13}
    V_i=V_i\cdot \one_{\{|V_i|\leq\varepsilon\}}+V_i\cdot \one_{\{|V_i|>\varepsilon\}}
  \end{equation} 
  and note, that $V_i \cdot \one_{\{|V_i| \leq \varepsilon\}}$ is bounded
  by $\|V_i \cdot \one_{\{|V_i| \leq \varepsilon\}}\| \leq \varepsilon$.
  As $V$ is non-negative, one has
  \begin{equation}
    \label{eq:1}
    \supp(V_i\cdot \one_{\{|V_i|>\varepsilon\}})\subseteq
    \bigcup_{j=1}^n\bigcup_{\substack{i=1\\i\not=j}}^n
  \left\{(x_1,\dots,x_n)\in\R^{nd}:\,V(x_i-x_j)\geq
    \frac{\varepsilon}{n(n-1)}\right\}. 
  \end{equation}
  As, by assumption (H.2), $V$ tends to $0$ at infinity,
  \eqref{eq:1}~implies that there exists $0<C(n;\varepsilon)$
  (independent of $L$) such that
  \begin{equation} 
    \label{gl4.15}
    \mu(\{|V_i| >\varepsilon \} \cap \Lambda_L^n) \leq C(n,\varepsilon)
    L^{(n-1)d}, 
  \end{equation} 
  where $\mu(\cdot)$ denotes the Lebesgue measure. Using decomposition
  (\ref{gl4.13}) of $V_i$, we obtain
  \begin{equation}
    \label{gl4.18}
    \begin{split}
      |\tr&(H_{n|\Lambda_L^n}+\lambda_0)^{-q}(H_{n|\Lambda_L^n}
      -z)^{-1} V_i(H_{|\Lambda_L^n}-z)^{-1} |\\
      &\leq \frac{\varepsilon}{|\Im z|^2}
      \tr|(H_{n|\Lambda_L^n}+\lambda_0)^{-q}| \\
      &\hskip3cm+\frac {1}{|\Im z|}
      \|V_i (H_{|\Lambda_L^n}-z)^{-1} \|\cdot
      \tr|(H_{n|\Lambda_L^n}+\lambda_0)^{-q} \one_{\{|V_i|
        > \varepsilon\} \cap \Lambda_L^n} |\\
      &\leq \frac{\varepsilon}{|\Im z|^2} \|(H_{n|\Lambda_L^n}
      +\lambda_0)^{-1} \|_{\cT_q}^q + \frac{1}{|\Im z|^2}
      \|(H_{n|\Lambda_L^n} +\lambda_0)^{-1}
      \|_{\cT_q}^{q-1} \\
      &\hskip4cm \cdot \| (H_{n|\Lambda_L^n} +\lambda_0)^{-1} \one_{\{|V_i|
        > \varepsilon\} \cap \Lambda_L^n} \|_{\cT_q}\cdot \|V_i
      (H_{n|\Lambda_L^n}+\lambda_0)^{-1}\|
    \end{split}
  \end{equation}
  where $\Vert\cdot\Vert_{\cT_q}$ denotes the $q$-th Schatten class
  norm (see~\cite{Si1}) and we used H{\"o}lder's inequality. In the same
  way, the cyclicity of the trace yields
  \begin{equation}
    \label{gl4.19}
    \begin{split}
      |\tr&(H_{n|\Lambda_L^n}+\lambda_0)^{l-q-1} V_i
      (H_{|\Lambda_L^n}+\lambda_0)^{-l}(H_{|\Lambda_L^n}-z)^{-1}|\\
      &\leq \tr |(H_{|\Lambda_L^n}+\lambda_0)^{-l}
      (H_{n|\Lambda_L^n}+\lambda_0)^{l-q-1} V_i
      (H_{|\Lambda_L^n}-z)^{-1}|\\
      &\leq \|(H_{|\Lambda_L^n}+\lambda_0)^{-l}
      (H_{n|\Lambda_L^n}+\lambda_0)^{l}\|
      \cdot \tr|(H_{n|\Lambda_L^n}+\lambda_0)^{-q-1} V_i
      (H_{|\Lambda_L^n}-z)^{-1}|\\
      &\leq \frac{C}{|\Im z|} \|(H_{n|\Lambda_L^n} +\lambda_0)^{-1}
      \|_{\cT_q}^{q-1} \cdot \|(H_{n|\Lambda_L^n} +\lambda_0)^{-1}
      \one_{\{|V_i| > \varepsilon\} \cap \Lambda_L^n} \|_{\cT_q}
      \cdot\|V_i (H_{n|\Lambda_L^n}+\lambda_0)^{-1}\|\\
      &\hskip8cm + C \frac{\varepsilon}{|\Im z|} \|
      (H_{n|\Lambda_L^n}+\lambda_0)^{-1}\|_{\cT_q}^q.
    \end{split}
  \end{equation}
  We are now left with estimating
  $\|(H_{n|\Lambda_L^n}+\lambda_0)^{-1}\|_{\cT_q}$ and
  $\|(H_{n|\Lambda_L^n}+\lambda_0)^{-1} \one_{\{|V_i| > \varepsilon\}
    \cap \Lambda_L^n}\|_{\cT_q}$ for $q$ sufficiently large, depending
  on $nd$. Therefore, we compute
  \begin{multline} 
     \label{gl4.20}
     \|(H_{n|\Lambda_L^n}+\lambda_0)^{-1} \one_{\{|V_i| >
       \varepsilon\} \cap \Lambda_L^n}\|_{\cT_q} \hspace{-0.1cm} \leq
     \|(H_{n|\Lambda_L^n}+\lambda_0)^{-1}
     (-\lap_{|\Lambda_L^n}+\lambda_0)^{\frac{1}{2}}\|_{\cT_{2q}}\\
     \cdot\|(-\lap_{|\Lambda_L^n}+\lambda_0)^{-\frac{1}{2}}
     \one_{\{|V_i| > \varepsilon\} \cap \Lambda_L^n}\|_{\cT_{2q}}
  \end{multline}
  We use the decomposition~\eqref{eq:2}. As the Laplacians are
  positive, the infinitesimal $-\Delta$-boundedness on $V_n^-$,
  Theorem X.18 of~\cite{RS2} and the definition of $\gamma$ imply the
  following form bound
  \begin{equation*}
    |\lkl\phi,V_{n|\Lambda_L^n}^-\phi\rkl|\leq\frac{1}{2}\lkl\phi,
    -\lap_{|\Lambda_L^n}\phi\rkl+\gamma\|\phi\|^2.
  \end{equation*}
  As $\lambda_0>2\gamma+1$, one has
  \begin{equation*}
    H_{n|\Lambda_L^n}+\lambda_0\geq-\lap_{|\Lambda_L^n}
    +V_{n|\Lambda_L^n}^-+\lambda_0
    \geq\frac12(-\lap_{|\Lambda_L^n}-2\gamma+2\lambda_0)
    \geq\frac12(-\lap_{|\Lambda_L^n}+\lambda_0).
  \end{equation*}
  Thus, the operator $H_{n|\Lambda_L^n}+\lambda_0$ is invertible and
  \begin{equation*}
    (H_{n|\Lambda_L^n}+\lambda_0)^{-1} \leq
    2(-\lap_{|\Lambda_L^n}+\lambda_0)^{-1}. 
  \end{equation*}
  Let $(\mu_{\ul{j}})_{\ul{j}}$ and $(\phi_{\ul{j}})_{\ul{j}}$
  respectively denote the eigenvalues and eigenfunctions of the
  Dirichlet Laplacian $-\lap_{|\Lambda_L^n}$ (the index ${\ul{j}}$
  runs over $(\N^{nd})^*$). For $q\in\N$ such that $2q>nd$, we compute
  \begin{equation*}
    \begin{split}
      \|(H_{n|\Lambda_L^n}+\lambda_0)^{-1}
      &(-\lap_{|\Lambda_L^n}+\lambda_0)^{\frac{1}{2}}\|_{\cT_{2q}}^{2q}
      =\sum_{\ul{j} \in \N^{nd}}
      (\mu_{\ul{j}}(-\lap_{|\Lambda_L^n})+\lambda_0)^q \lkl
      \phi_{\ul{j}},
      (H_{n|\Lambda_L^n}+\lambda_0)^{-1} \phi_{\ul{j}} \rkl^{2q}\\
      &\leq 2^{2q}\sum_{\ul{j} \in \N^{nd}}
      (\mu_{\ul{j}}(-\lap_{|\Lambda_L^n})+\lambda_0)^q \lkl
      \phi_{\ul{j}},(-\lap_{|\Lambda_L^n}+\lambda_0)^{-1}
      \phi_{\ul{j}} \rkl)^{2q} \\
      &= 2^{2q}\sum_{\ul{j} \in \N^{nd}} (\mu_{\ul{j}}
      (-\lap_{|\Lambda_L^n})+\lambda_0)^{-q}\leq CL^{nd}.
    \end{split}
  \end{equation*}
  The last estimate is a direct computation using the explicit
  form of the Dirichlet eigenvalues.\\
  By Lemma 2.2 in~\cite{KlPa}, we know that, for $q\in\N$ such that
  $2q>nd$, there exists $C_q>0$ such that, for any measurable subset
  $\Lambda'\subseteq\Lambda_L^n$, one has
  \begin{equation} 
    \label{gl4.16}
    \|(-\lap_{|\Lambda_L^n}+\lambda_0)^{-\frac{1}{2}}
    \one_{\Lambda'}\|_{\cT_{2q}}^{2q} \leq C_q \mu(\Lambda').
  \end{equation} 
  Choosing $\Lambda'=\{|V_i|>\varepsilon\} \cap \Lambda_L^n$ and
  taking (\ref{gl4.15}) into account, then by combining estimates
  (\ref{gl4.18})--(\ref{gl4.16}), we get that there exists $c$,
  depending only on $q$ (and the bound in assumption (H.3)), such that
  \begin{multline*}
    \tr| (H_{|\Lambda_L^n}+\lambda_0)^{-q} (H_{|\Lambda_L^n}-z)^{-1}-
    (H_{n|\Lambda_L^n}+\lambda_0)^{-q} (H_{n|\Lambda_L^n}-z)^{-1}| \\ 
    \leq c(\frac{\varepsilon}{|\Im z|^2} L^{nd}+ \frac{1}{|\Im z|^2}
    L^{nd-\frac{d}{2q}}+ \frac{\varepsilon}{|\Im z|} L^{nd}+
    \frac{1}{|\Im z|} L^{nd-\frac{d}{2q}}).
  \end{multline*}
  Using this inequality in (\ref{gl4.12}), we get (\ref{gl4.11}) as,
  $\tilde{\varphi}$ being almost analytic,
  $\overline{\partial}\tilde{\varphi}(z)$ vanishes to any order in
  $\Im z$ as $z$ approaches the real line.  Thus, we completed the
  proof of Theorem~\ref{th4.2}.\qed

\begin {thebibliography}{999999}
\bibitem [CHK] {CHK} Jean-Michel Combes, Peter Hislop, Fr{\'e}d{\'e}ric Klopp:
  Local and Global Continuity of the Integrated Density of States;
  Contemporary Mathematics; {\bf 327}, 61-74 (2003)
\bibitem [Kl] {Kl} Fr{\'e}d{\'e}ric Klopp: Internal Lifshits tails for random
  perturbations of periodic Schr{\"o}dinger operators; Duke Mathematical
  Journal; {\bf 98}(2) 335-396, (1999)
\bibitem [KlPa] {KlPa} Fr{\'e}d{\'e}ric Klopp, Leonid Pastur: Lifshitz tails
  for random Schr{\"o}dinger operators with negative Poisson potential;
  Communications in Mathematical Physics; {\bf 206}, 57-103 (1999)
\bibitem [PaFi] {PaFi} Leonid Pastur, Alexander Figotin: Spectra of
  Random and Almost-Periodic Operators; Springer Verlag (1992)
\bibitem [RS2] {RS2} Michael Reed, Barry Simon: Methods of Modern
  Mathematical Physics; Volume II: Fourier Analysis, Self-Adjointness,
  Academic Press, San Diego 1975
\bibitem [RS4] {RS4} Michael Reed, Barry Simon: Methods of Modern
  Mathematical Physics; Volume IV: Analysis of Operators, Academic
  Press, San Diego 1978
\bibitem [Si]{Si1} Barry Simon: Trace ideals and their application;
  Cambridge University Press, Cambridge 1979
\bibitem [Stl]{Stl} Peter Stollman: Caught by disorder; Birkh{\"a}user
  2001
\bibitem [Stz] {Stz} G{\"u}nter Stolz: Strategies in localization proofs
  for one-dimensional random Schr{\"o}dinger operators; Proc. Indian Acad.
  Sci. (Math. Sci.); {\bf 112}, 229-243 (2002)
\bibitem [Ze]{Ze2} Heribert Zenk: An interacting multielectron
  Anderson model; Preprint mp\_arc 03-410
\end{thebibliography}
\end{document}